\documentclass[twocolumn,prl,floatfix,citeautoscript,nofootinbib,superscriptaddress]{revtex4}
\usepackage{amsbsy}
\usepackage{latexsym,epsfig,graphicx}
\usepackage{dcolumn}
\usepackage{graphicx}
\usepackage{subfigure}
\usepackage{comment}
\usepackage{color}
\usepackage{bm}
\usepackage{mathrsfs}
\usepackage{amsfonts}
\usepackage{amsmath}
\usepackage{color}
\usepackage{amssymb}
\usepackage{xspace}
\usepackage{epstopdf}
\usepackage{tabularx}
\usepackage{longtable}
\usepackage[colorlinks=true, letterpaper=true, pdfstartview=FitV, linkcolor=blue, citecolor=blue, urlcolor=blue]{hyperref}
\usepackage[normalem]{ulem}

\pdfoutput=1

\begin{document}

\title{Tunable spin-orbit coupling and magnetic superstripe phase in a BEC}
\author{Xi-Wang Luo}
\author{Chuanwei Zhang}
\thanks{Corresponding author. \\
Email: \href{mailto:chuanwei.zhang@utdallas.edu}{chuanwei.zhang@utdallas.edu}%
}
\affiliation{Department of Physics, The University of Texas at Dallas, Richardson, Texas
75080-3021, USA}

\begin{abstract}
Superstripe phases in Bose-Einstein condensates (BECs),
possessing both crystalline structure and superfluidity, opens a new
avenue for exploring exotic quantum matters---supersolids.
However, conclusive
detection and further exploration of a superstripe is still challenging in experiments because of
its short period, low visibility, fragility against
magnetic field fluctuation or short lifetime. Here we propose a scheme in
a spin-orbit coupled BEC which overcomes these obstacles and
generates a robust magnetic superstripe phase,
with only spin (no total) density modulation due to the magnetic
translational symmetry,
ready for direct real-space observation.
In the scheme, two hyperfine spin states are individually Raman coupled with
a largely-detuned third state, which induce a momentum-space separation
between two lower band dispersions, yielding an effective spin-1/2 system
with tunable spin-orbit coupling and Zeeman fields. Without effective Zeeman
fields, spin-dependent interaction dominates, yielding a magnetic superstripe
phase with a long tunable period and high visibility. Our scheme provides a
platform for observing and exploring exotic properties of superstripe phases
as well as novel physics with tunable spin-orbit coupling.
\end{abstract}

\maketitle

\emph{Introduction.---} In supersolids, crystalline and superfluidity orders
are formed through spontaneously breaking continuous translational and U(1)
gauge symmetries~\cite{boninsegni2012colloquium}. The concept of
supersolidity was originally discussed in solid $^{4}$He~\cite%
{thouless1969the, andreev1971quantum}, and later generalized to other
superfluid systems that spontaneously form spatial periodicity. In
particular, ultracold atomic gases provide a powerful platform for exploring
quantum phases with supersolid-like properties \cite{cinti2014defect, baumann2010dicke,
PhysRevLett.113.070404, leonard2017supersolid, PhysRevLett.122.130405,PhysRevX.9.011051,PhysRevX.9.021012}. For instance, a superstripe
phase with spontaneously formed periodic density modulation has been
theoretically proposed for a spin-orbit (SO) coupled Bose-Einstein
condensate (BEC) with anisotropic spin interactions \cite{li2012quantum, zhang2012mean, ho2011bose, wang2010spin, PhysRevLett.110.235302}. In this context, the
recent experimental realization of SO coupling in ultracold atoms \cite%
{lin2011spin, zhang2012collective, qu2013observation, olson2014tunable, ji2014experimental,
wang2012spin, cheuk2012spin, Williams2013, wu2016realization, huang2016experimental,
campbell2015itinerant, luo2016tunable} paves a promising path for the
observation and exploration of the long-sought supersolid phases. Here the pseudospin states could
be formed by either two atomic hyperfine ground states~\cite%
{Stanescu2008, Wu2011, hu2012spin, ozawa2012stability,
galitski2013spin, lan2014raman, Natu2015} or two sites of a double well
optical lattice \cite{PhysRevLett.117.185301}. For the later case,
the crystalline structures of a BEC have been indirectly observed recently using Bragg
reflection \cite{li2017stripe}.

There are a few major obstacles~\cite{martone2014approach, ho1998spinor,
Ohmi1998Bose} for conclusive
observation
and further exploration of superstripe phases in a
SO coupled BEC: \textit{i}) A superstripe is formed by the superposition of
two plane waves separated by a large momentum, leading to a short period at
the order of optical wavelength~for the density modulation~\cite%
{lin2011spin, zhang2012collective}; \textit{ii}) A superstripe phase is
energetically unfavorable by density interaction $g_{0}$ due to its total
density modulation, therefore could only exhibit a low visibility and exist
in a small parameter region favored by weak spin interaction~\cite%
{li2012quantum, ho2011bose}; \textit{iii}) The superstripe phase for
hyperfine state pseudospins is fragile against magnetic field fluctuation
because the relative energy between two spin states is sensitive to the
magnetic field~\cite{qu2013observation, zhang2012mean, ji2014experimental};
\textit{iv}) The superstripe phases for
double-well lattice pseudospins (where SO coupling is realize by additional moving lattices)
or dipole gases
have a short lifetime~\cite{li2017stripe,PhysRevLett.122.130405,PhysRevX.9.011051,PhysRevX.9.021012}.

In this paper, we propose that all these obstacles can be completely
overcome by engineering an effective spin-1/2 subsystem with tunable SO
coupling in a spin-1 BEC
(we use atomic hyperfine-state pseudospins to avoid heatings)~\cite{sun2016interacting, yu2016phase,
martone2016tricriticalities, PhysRevLett.119.193001}, leading to
a promising scheme for in-depth investigation of
supersolidity. Our main results are:

\textit{i}) We propose a generic and experimentally feasible scheme for
generating an effective spin-1/2 system with tunable SO coupling through two
individual Raman couplings of two spin states ($\left\vert \uparrow
\right\rangle $, $\left\vert \downarrow \right\rangle $) with a third higher
energy state ($\left\vert 0\right\rangle $), which induce a momentum
separation between two lower band dispersions, yielding SO coupling. The SO
coupling strength can be widely tuned by varying laser and microwave
intensities, in contrast to fixed SO coupling strength determined by the
laser geometry in previous experiments \cite{lin2011spin}.

\textit{ii}) Because the SO coupling
is induced by the Raman coupling with the third state, it can exist without an
effective transverse field, where the total density modulation vanishes (due to magnetic translational symmetry)
even when both band minima are occupied by the BEC.
In this case, the spin
interaction $g_{2}$, instead of density interaction $g_{0}$, dominates the
phase diagram, leading to novel high-visibility ($\sim$100\%) magnetic superstripe phase
with only spin density modulation.
Depending on the SO coupling strength, the superstripe period is
tunable up to $\sim 5\mu m$, which can be directly imaged in the real space.
Finally, the relative
energy between two band minima is insensitive to magnetic field
fluctuations, making the superstripe phase robust in experiments.

\textit{iii}) Beside superstripe phases, we find a rich phase diagram with
other novel phases in different parameter regions.


\begin{figure}[t]
\includegraphics[width=1.0\linewidth]{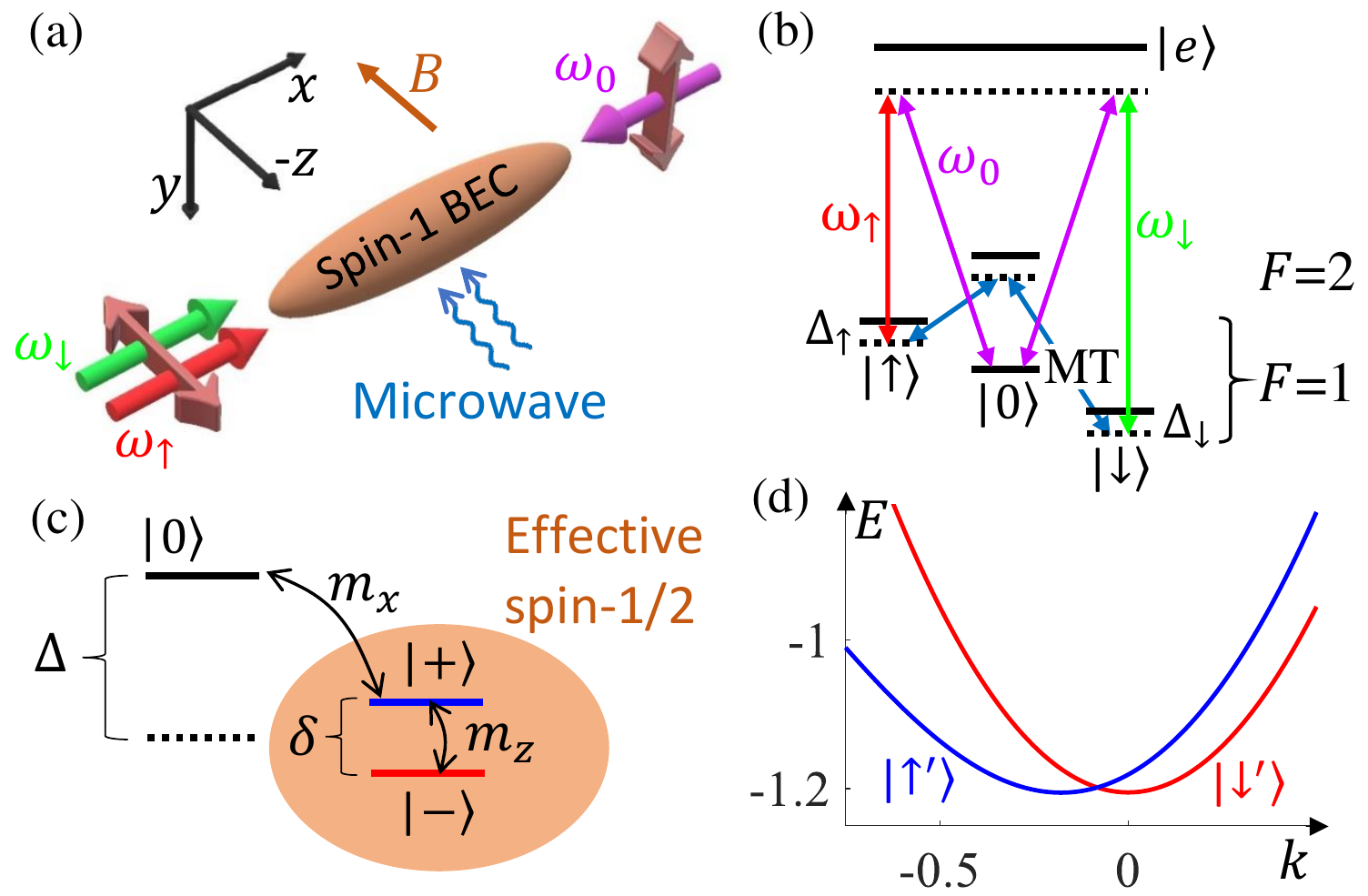}
\caption{(a) Experimental scheme to generate SO coupling for an effective
spin-1/2 system using a spin-1 BEC. The bias field $B$ is along the $z$
direction. (b) The corresponding two-photon Raman transitions and microwave
transitions (MT) between three hyperfine spin states. (c) Mapping to an
effective spin-1/2 system, with spin states $\left\vert \pm \right\rangle =%
\frac{1}{\protect\sqrt{2}}(\left\vert \uparrow \right\rangle \pm \left\vert
\downarrow \right\rangle )$, $\Delta =(\Delta _{\uparrow }+\Delta
_{\downarrow })/2$, and $m_{z}=(\Delta _{\uparrow }-\Delta _{\downarrow })/2$%
. (d) Two lower bands for $m_{x}=1$, $m_{z}=0$, $\protect\delta =0.205$, and
$\Delta =-1$.}
\label{fig:sys}
\end{figure}

\emph{Experimental scheme and Hamiltonian:---} We consider an experimental
setup shown in Fig.~\ref{fig:sys}(a), which is similar as that in a recent
experiment~\cite{campbell2015itinerant} but with different laser
configuration and additional microwave fields. Three Raman lasers are
employed to couple hyperfine states $\left\vert \uparrow ,\downarrow
\right\rangle $ with $\left\vert 0\right\rangle $ in the $F=1$ manifold of $%
^{87}$Rb atoms [see Fig.~\ref{fig:sys}(b)], with $2k_{\text{R}}$ momentum
transfer. $\left\vert \downarrow \right\rangle $ and $\left\vert \uparrow
\right\rangle $ are coupled by a two-photon microwave transition via an
intermediate virtual state $|F=2,m_{F}=0\rangle $~\cite%
{PhysRevLett.81.243, mertes2007nonequilibrium, PhysRevA.87.053614} with zero
momentum transfer. 
After a unitary transformation $U=\exp (i2k_{\text{R}}x)|0\rangle \langle 0|
$ that only transforms state $|0\rangle $ to a quasi-momentum basis, the
resulting single-particle Hamiltonian becomes 
\begin{eqnarray}
H_{0} &=&\hat{k}^{2}-(4\hat{k}+4)(F_{z}^{2}-1)+\Delta F_{z}^{2}  \notag \\
&&+\sqrt{2}m_{x}F_{x}+m_{z}F_{z}+\delta (F_{x}^{2}-F_{y}^{2}).  \label{eq:H0}
\end{eqnarray}%
Here we set $\hbar =1$ and use the energy and momentum units $\frac{k_{R}^{2}%
}{2m}$ and $k_{R}$. $F_{i}$ ($i=x,y,z$) are spin vectors and $4\hat{k}%
F_{z}^{2}$ describes the spin-tensor-momentum coupling \cite%
{PhysRevLett.119.193001}. $m_{x}$ ($\delta $) is the Raman (microwave)
coupling strength between $|0\rangle $ and $\left\vert \uparrow ,\downarrow
\right\rangle $ ($\left\vert \uparrow \right\rangle $ and $\left\vert
\downarrow \right\rangle $), which can be tuned with high precision. The
phase difference between two Raman lasers with frequencies $\omega
_{\uparrow }$ and $\omega _{\downarrow }$ is locked to the same value as
that between two microwave fields such that $m_{x}$ and $\delta $ become
real and positive by gauging out irrelevant phases. $m_{z}$
and $\Delta $ 
are linear and quadratic Zeeman fields that can be tuned by laser detunings.

\emph{Tunable SO coupling strength.---}We consider 
a large $\Delta \ll 0$ such that low energy dynamics are mainly
characterized by spin states $\left\vert \uparrow \right\rangle $ and $%
\left\vert \downarrow \right\rangle $ with two band minima near $k=0$ [see
Figs.~\ref{fig:sys}(c) and (d)].
By hybridizing $\left\vert +\right\rangle $ [$\left\vert \pm \right\rangle
\equiv \frac{1}{\sqrt{2}}(\left\vert \uparrow \right\rangle \pm \left\vert
\downarrow \right\rangle )$] with $\left\vert 0\right\rangle $ [see Fig.~\ref%
{fig:sys}(c)] for state $\left\vert \uparrow ^{\prime }\right\rangle $, the
Raman coupling $m_{x}$ induces a momentum shift for $\left\vert \uparrow
^{\prime }\right\rangle $ band with band minimum $k_{\text{m}}<0$ [see Fig.~%
\ref{fig:sys}(d)]. The band for $\left\vert \downarrow ^{\prime
}\right\rangle \equiv \left\vert -\right\rangle $ is unaffected by $m_{x}$.
To restore the degeneracy between two band minima, a two-photon microwave
transition with $\delta >0$ is used to tune their relative energy, forming
an effective spin-1/2 system [see Fig.~\ref{fig:sys}(d)]. Here $\delta $ is
crucial because $\left\vert \downarrow ^{\prime }\right\rangle $ band would
be always higher than $\left\vert \uparrow ^{\prime }\right\rangle $
band~without $\delta $ \cite{PhysRevLett.119.193001}.

The low energy effective Hamiltonian in the basis $\{\left\vert \uparrow
^{\prime }\right\rangle ,\left\vert \downarrow ^{\prime }\right\rangle \}$
can be written as
\begin{equation}
H_{\text{eff}}=\left[
\begin{array}{cc}
\eta (k-k_{\text{m}})^{2} & 0 \\
0 & k^{2}%
\end{array}%
\right] +B_{z}\sigma _{z}+B_{x}\sigma _{x},
\end{equation}%
leading to a SO coupling $\eta k_{\text{m}}k\sigma _{z}$. The effective
\textquotedblleft detuning\textquotedblright\ $B_{z}$ and \textquotedblleft
Raman coupling\textquotedblright\ $B_{x}$ between $\left\vert \uparrow
^{\prime }\right\rangle $ and $\left\vert \downarrow ^{\prime }\right\rangle
$ bands can be tuned by $\delta $ and $m_{z}$ respectively (see Appendix). $\eta $
is the mass ratio between $\left\vert \uparrow ^{\prime }\right\rangle $ and
$\left\vert \downarrow ^{\prime }\right\rangle $ and $k_{\text{m}}$
characterizes the SO coupling strength, which can be tuned by varying Raman
laser intensities (i.e., $m_{x}$). In contrast, the SO strength is preset by
Raman laser geometry \cite{galitski2013spin} in previous experiments and its
modulation through periodic fast modulation of laser intensities \cite%
{zhang2013tunable, jimenez2015tunable, grusdt2017tunable} may lead to significant heating
issues and complex interaction effects.

Our scheme for tunable 1D SO coupling only relies on the existence of three
hyperfine ground states that can be coupled with each other, therefore it
can be applied to other alkali (e.g., potassium) and alkaline-earth(-like)
atoms (e.g., strontium, ytterbium). The corresponding laser
configurations could be slightly different (see Appendix).

\emph{Interacting phase diagram.---} In the presence of atomic interaction,
the effective spin-1/2 system with tunable SO coupling provides a path for
realize superstripe phases with long period and high visibility. For the
simplicity of the presentation and accurate description of the results, we,
however, still use the original spin-1 Hamiltonian (\ref{eq:H0}) for our
calculation.

The interaction energy density can be expressed as (see Appendix)
\begin{equation}
\varepsilon _{\text{int}}\ =\frac{1}{V}\int dx\left[ \frac{g_{0}}{2}%
n_{\text{tot}}^{2}+g_{2}n_{0}(n_{\uparrow }+n_{\downarrow})+\frac{g_{2}}{2}\mathcal{F}_{z}^{2}\right] ,  \label{eq:inter-energydensity}
\end{equation}%
where $V$ is the system volume and $n_\text{tot}$, $n_{i}$ ($i=0,\uparrow ,\downarrow $%
) are the total and spin densities, with $\mathcal{F}_{z}\equiv n_{\uparrow }-n_{\downarrow}$
the polarization and $g_{0}$, $g_{2}$ the
density- and spin-interaction strengths.
Under the Gross-Pitaevskii (GP) approximation, we adopt a variational ansatz
as the general superposition of two plane waves around two band minima
\begin{equation}
\Psi =\sqrt{\bar{n}}\left( |c_{1}|\chi _{1}e^{ik_{1}x}+|c_{2}|\chi
_{2}e^{ik_{2}x+i\alpha }\right) ,  \label{eq:ans}
\end{equation}%
which is normalized by the average particle number density $\bar{n}%
=V^{-1}\int dx\Psi ^{\dag }\Psi $, with three-component spinors $\chi
_{j}=(\cos \theta _{j}\cos \phi _{j},-\sin \theta _{j},\cos \theta _{j}\sin
\phi _{j})^{T}$ and $|c_{1}|^{2}+|c_{2}|^{2}=1$. The ground state is
determined by minimizing the total energy density
\begin{equation}
\varepsilon _{\text{tot}}\ =\varepsilon _{\text{int}}+\frac{1}{V}\int dx\Psi
^{\dag }H_{0}\Psi   \label{eq:tot-energydensity}
\end{equation}%
with respect to eight variational parameters $|c_{1}|$, $k_{1}$, $k_{2}$, $%
\theta _{1}$, $\theta _{2}$, $\phi _{1}$, $\phi _{2}$, and $\alpha $ (see Appendix). The phase diagram can be characterized by the atomic total
density $n_{\text{tot}}$, spin density $n_{i}$ and polarization $\langle
F_{z}\rangle $ which can be measured directly in experiments.
We also obtain the ground states by directly simulating GP equation
numerically, which are in good agreement with the variational results.

\begin{figure}[t]
\includegraphics[width=1.0\linewidth]{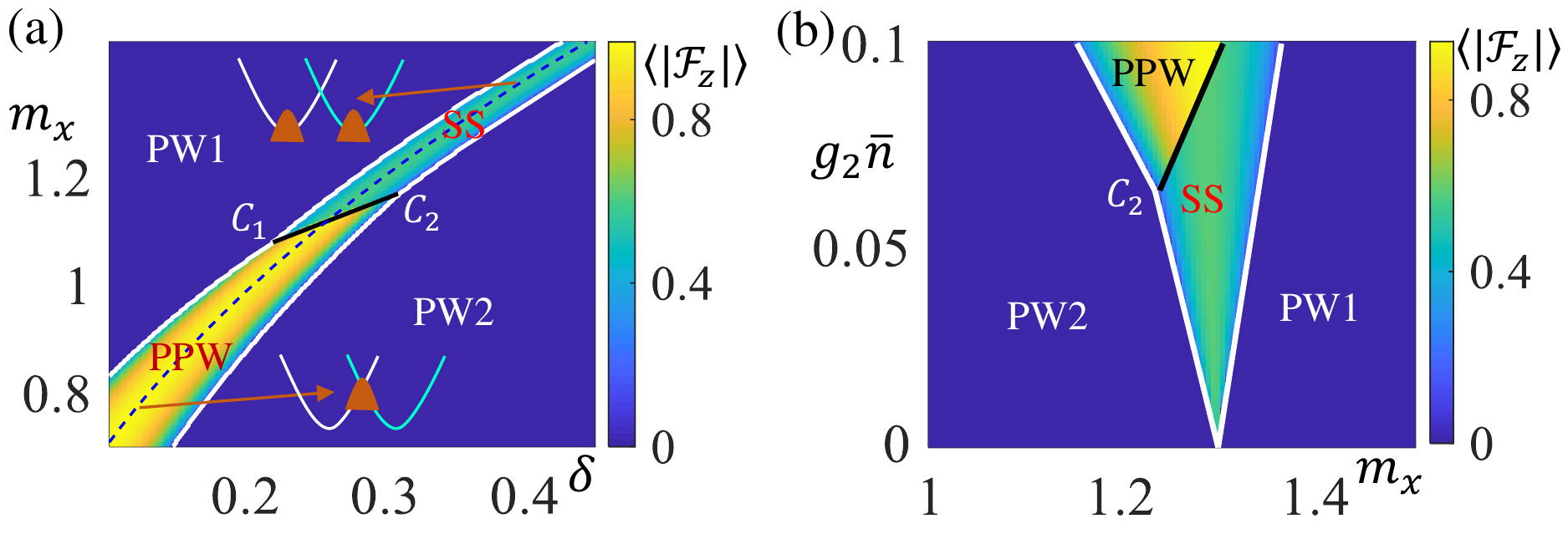}
\caption{(a) Phase diagram in the $m_{x}$-$\protect\delta $ plane with $g_{2}%
\bar{n}=-0.05$, $\Delta =-1$, and $m_{z}=0$. Color bar shows the property of
the polarization density (average of its absolute value). (b) Phase diagram
in the $m_{x}$-$g_{2}$ plane with $\protect\delta =0.35$, other parameters
are the same as in (a). Black (white) solid lines correspond to first (second) order
phase transitions.}
\label{fig:phase}
\end{figure}


We first consider $m_{z}=0$, where the spin states of two lower bands
are orthogonal (i.e., $\langle\chi _{1}|\chi _{2}\rangle=0$
for $B_{x}=0$). Therefore, the total density is always
a constant, and the density interaction $g_{0}$ plays no role for the phase
diagram. 
The spin interaction $g_{2}$ tends to lower the
energy by occupying both band minima, leading to a superstripe
ground state. The phase diagram obtained from the variational method for
ferromagnetic spin interaction (e.g., $^{87}$Rb with $g_{2}<0$) is shown in
Fig.~\ref{fig:phase}(a) as a function of Raman couplings $m_{x}$ and $\delta
$. There are four phases: the plane-wave phase PW1 (PW2) with zero spin
polarization (i.e., $\mathcal{F}_{z}=0$) and
single momentum occupation at the left (right) band minimum; the polarized
plane-wave phase PPW with uniform spin polarization (i.e., $\mathcal{F}_{z}\neq 0$%
) and single momentum occupation at the barrier between two band minima; the
magnetic superstripe phase SS with striped spin polarizations $\mathcal{F}_{z}$ (total density is uniform)
and momentum occupations at both band minima [see the inset in Fig.~\ref%
{fig:phase}(a)].
The plane-wave phases preserves the continuous translational symmetry with
$\hat{T}_d|\Psi\rangle=e^{ik_1d} |\Psi\rangle$, where $\hat{T}_d$ is
the translation operator. For the
SS phase, we have $\hat{T}_d|\Psi\rangle=\Lambda_d |\Psi\rangle$ with
$\Lambda_d$ a spatial-independent
unitary matrix because of $\langle\chi _{1}|\chi _{2}\rangle=0$.
In particular, we have $\Lambda_d=e^{ik_1d} |\chi _{1}\rangle\langle\chi _{1}|+e^{ik_2d} |\chi _{2}\rangle\langle\chi _{2}|$.
This means that
the SS phase breaks the translational symmetry but
preserves a magnetic translational symmetry $\hat{T}_m|\Psi\rangle=|\Psi\rangle$ with
$\hat{T}_m=\Lambda_d^\dag\hat{T}_d$. This magnetic translational symmetry
is responsible to the uniform total density [since $n_\text{tot}(x+d)=|\hat{T}_m\Psi|^2=|\Psi|^2=n_\text{tot}(x)$].

Both PPW and SS phases result from the ferromagnetic spin interaction, and
the total energy is minimized by generating non-zero spin polarizations $%
\mathcal{F}_{z}$ (uniform in PPW and striped in SS).
We note that only state $|0\rangle $ is transformed to the quasi-momentum
basis, therefore the spin density modulation $\mathcal{F}_{z}$ in SS phase
are unaffected after transforming back to the real mechanical momentum.
The uniform polarization $\mathcal{F}_{z}$ in PPW phase can be either
positive or negative due to the spontaneously breaking of the discrete $Z_{2}
$ symmetry between states $\left\vert \uparrow \right\rangle $ and $%
\left\vert \downarrow \right\rangle $. In the supersolid-ordered SS phase, $%
n_{\uparrow }$ and $n_{\downarrow }$ exhibit out-of-phase density
modulations (therefore leading to a nonzero spin-polarization modulation $%
\mathcal{F}_{z}$) that spontaneously break the continuous translational
symmetry due to the arbitrariness of relative phase $\alpha $ between two $k$
states.
We can always choose the relative strength between the Raman and
microwave couplings such that two band minima are degenerate [see the dashed
line in Fig.~\ref{fig:phase}(a)]; therefore, the SS phase can exist in a
long ribbon along the degenerate line in the $m_{x}$-$\delta $ plane.

\begin{figure}[t]
\includegraphics[width=1.0\linewidth]{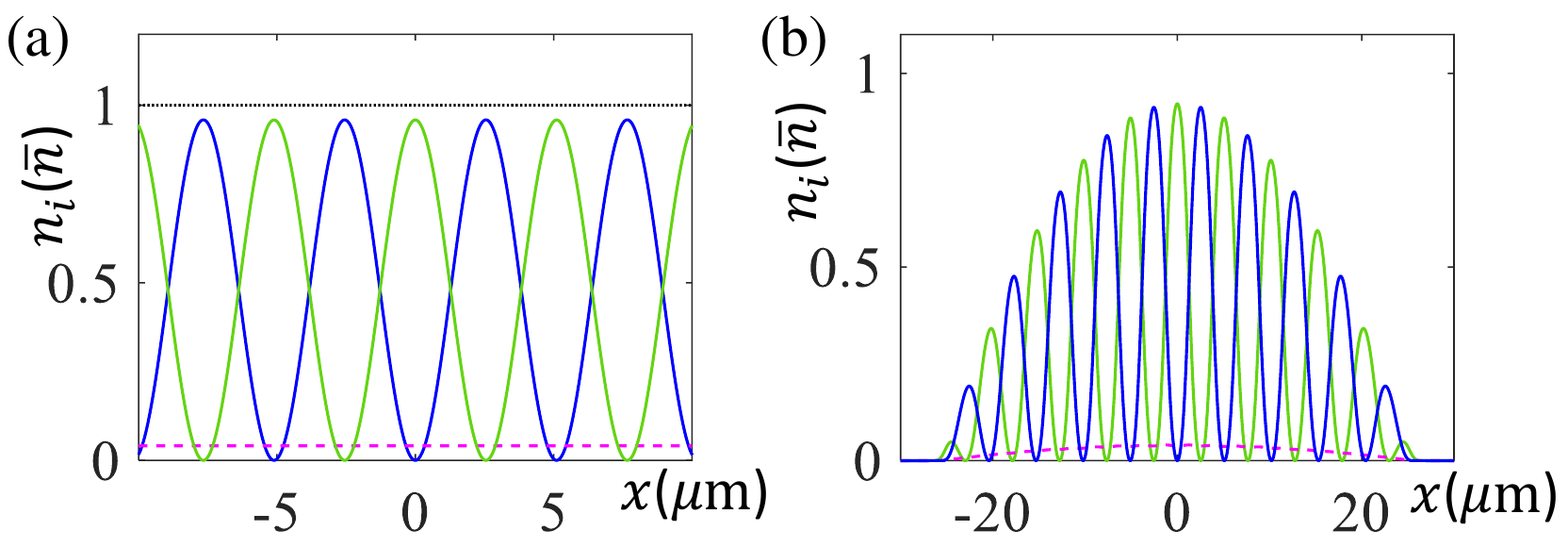}
\caption{Spin density modulations in the SS phase. (a) Ground state obtained
from the variational ansatz for a non-trapped BEC. (b) Ground state for a
trapped BEC (with trapping frequency 50Hz) obtained by directly solving the GP equation.
Common parameters: $g_{2}\bar{n}=-0.01$, $%
g_{0}=200|g_{2}|$, $\protect\delta =0.18$, and $m_{x}=0.938$ ($m_{x}$ is
chosen to obtain two degenerate band minima), and other parameters are the same
as in Fig.~\protect\ref{fig:phase}(a). blue (black) solid, green (light gray) solid, black dotted and purple dashed lines
correspond to $n_{\uparrow }$, $n_{\downarrow }$,
$n_{0}$ and $n_{\text{tot}}$, respectively.
Raman lasers with 790nm wavelength (typical for
alkali atoms) are used.}
\label{fig:SS_density}
\end{figure}

For a strong Raman coupling $m_{x}$, where two band minima are well
separated [the upper part in Fig.~\ref{fig:phase}(a)], the ground state
prefers a plane wave (PW1) at the left band minimum when the microwave
transition $\delta $ is weak.
As we increase $\delta $ [which would rise (lower) the left (right) band
minimum], 
the BEC starts to partially occupy the right band minimum, undergoing a
second-order phase transition to the magnetic superstripe phase (SS) where both
minima are populated. By further increasing $\delta $, the population of the
right (left) minimum increases (decreases) until another second-order phase
transition occurs where the BEC is fully transferred to the low-energy right
minimum (PW2).

For weak Raman coupling $m_{x}$ 
[the lower part of the diagram of Fig.~\ref{fig:phase}(a)], the two band
minima are too close in momentum space to form the magnetic superstripe phase.
If the system starts at the PW1 phase, 
it undergoes a second-order phase transition to PPW phase as
$\delta $ increases, where the BEC would not partially occupy the right band
minimum, but instead, starts to occupy two lower bands at the same momentum,
generating a uniform spin polarization. Therefore, the BEC stays in a
plane-wave state and shifts towards the right band minimum as a whole, where
a second-order phase transition to PW2 occurs. The transition between SS and
PPW phase is of first order, with their phase boundary ending at two triple
points $C_{1,2}$, as shown in Fig.~\ref{fig:phase}(a).
Compared with the SS phase, the PPW phase has a higher single-particle
energy, but the total energy is favorable due to lower spin-interaction
energy from its uniform spin polarization.
As a result, 
the system prefers the PPW phase for weak Raman coupling $m_{x}$ where the
SO coupling is weak and band barrier is low. 
For conventional SO coupled spin-1/2 systems, atoms may condense at the
barrier maximum only for very strong Raman coupling or interaction~\cite%
{li2012quantum}. 

In Fig.~\ref{fig:phase}(b) we plot the phase diagram in the $g_{2}$-$m_{x}$
plane with a fixed $\delta $. We see that the areas of PPW and SS phases
shrink as $|g_{2}|$ decreases. 
and the PPW phase in the weak SO coupling region are replaced by the
SS phase as $g_{2}$ decreases. Therefore, the SS phase can have even longer period for
weaker spin interaction.
In Fig.~\ref{fig:phase}(b) with spin interaction $g_{2}\bar{n}=-0.05$, the
superstripe period can be up to around 3.8$\mu $m (see Appendix). For
$g_{2}\bar{n}=-0.01$, the period can be greater than 5$\mu $%
m, as shown in Fig.~\ref{fig:SS_density}(a).
Due to the uniform total density ($n_{0}$ is also uniform),
the density interaction
$g_{0}$ is irrelevant, and
the spin interaction $g_{2}$ can lead to high-visibility ($\sim$100\%) spin modulations in the SS phase, where
the spin densities $n_{\uparrow ,\downarrow }$ show out-of-phase modulation with
a long period and high visibility. 
In Fig.~\ref{fig:SS_density}(b) we show the density distributions in the presence
of a realistic harmonic trap, which are obtained by numerical simulation of
the GP equation directly. Such long-period ($\sim 5\mu $m) and
high-visibility ($\sim100\%$) magnetic superstripes can be directly detected by
real-space imaging~\cite{leblanc2013direct, leder2016real, parsons2015site}.
We emphasize that, here the long-period, high-visibility superstripe phase
is the ground state possessing true supersolidity, which is different from
the dynamically generated excited superstripe state~\cite%
{PhysRevLett.119.193001}. 

\begin{figure}[t]
\includegraphics[width=1.0\linewidth]{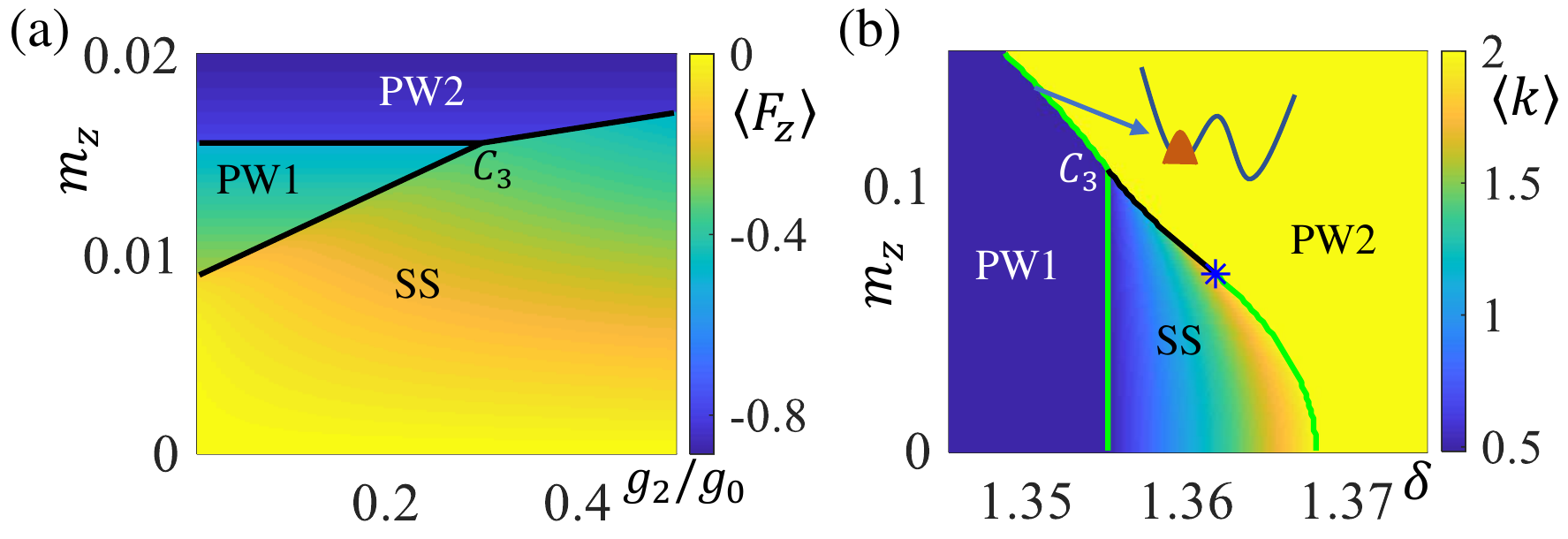}
\caption{(a) Phase diagram in the $m_{z}$-$g_{2}/g_{0}$ plane with $%
m_{x}=1.2 $, $\Delta =-1$ and $g_{2}\bar{n}=-0.01$. Color bar shows the
averaged polarization $\langle F_{z}\rangle $. Though the SS phase can exist
in a large interval of $\protect\delta $, here PW1 is sensitive to $\protect%
\delta $ and we use $\protect\delta =0.298$ to make the PW1 region large.
(b) Phase diagram in the $m_{z}$-$\protect\delta $ plane with $m_{x}=2$, $%
\Delta =0$, $g_{2}\bar{n}=-0.01$ and $g_{0}=200|g_{2}|$. Color bar shows
averaged momentum. The SS-PW2 transition changes from second order [green (light gray)
line] to first order (black line) at the point marked by a star.}
\label{fig:mz_g0}
\end{figure}

\emph{Zeeman field effects.---} So far we have focused on the case with zero
linear Zeeman field $m_{z}=0$. In a realistic experiment, though the
detunings of laser frequencies can be tuned with high accuracy, the magnetic
field fluctuation would lead to a non-zero $m_{z}$. Therefore, the
robustness of the superstripe phase against Zeeman field fluctuation is very
important. 
In a conventional SO coupled spin-1/2 system~\cite{lin2011spin,
ji2014experimental}, the spin states are represented directly by the
hyperfine states, leading to two band minima whose energies are sensitive to
magnetic fields~\cite{li2012quantum}. The superstripe phase is stable only
in a narrow width $|m_{z}|\lesssim g_{2}\bar{n}/4$, which requires extreme
control of ambient magnetic field fluctuations that is very challenging~\cite%
{li2017stripe}. In our system,
$m_{z}$ acts like an effective \textquotedblleft Raman coupling" which opens
a band gap at the crossing point between two lower bands.
We find that the SS phase could be very robust against such effective
\textquotedblleft Raman coupling".

In the presence of $m_{z}$, the spin state at the two band minima are no
longer orthogonal, and the SS phase now possesses both spin and total
density modulations (see Appendix), where $g_{0}$ becomes important and favors
the plane-wave phases at large $|m_{z}|$.
In Fig.~\ref{fig:mz_g0}(a), we plot the phase diagram in the $m_{z}$-$%
g_{2}/g_{0}$ plane with fixed $g_{2}$,
and a small $m_{x}$ is used 
to obtain a small SO coupling $k_{\text{m}}\sim k_{R}/4$ (corresponding to a
long superstripe period $\sim 3.2\mu $m enough for direct real-space
observation~\cite{leblanc2013direct, leder2016real, parsons2015site}). We
find that even for strong density interaction $|g_{2}|/g_{0}\sim 0.005$
(typical for $^{87}$Rb atoms), the long-period, high visibility
superstripes can exist up to a large Zeeman field $|m_{z}|\sim g_{2}\bar{n}$
without involving strong total-density modulations.

The SS phase becomes more robust against $m_{z}$ in the strong SO coupling
region. Fig.~\ref{fig:mz_g0}(b) shows the phase diagram in the $m_{z}$-$%
\delta $ for $k_{\text{m}}\sim 1.5k_{\text{R}}$ (corresponding to a short
superstripe period which may be observed by Bragg reflection). The system
may stay in the SS phase until it shrinks to the triple point $C_{3}$ at
extremely strong Zeeman field $m_{z}\sim 10g_{2}\bar{n}$. The transition
order can be revealed by looking at the behavior of $\langle k\rangle $, $%
\langle F_{z}\rangle $ or the visibility (i.e., a jump in $\langle k\rangle $%
, $\langle F_{z}\rangle $ or visibility represents a first-order
transition). It is worth to mention that for $m_{z}\neq 0$, the Hamiltonian
no longer has the symmetry between $\left\vert \uparrow \right\rangle $ and $%
\left\vert \downarrow \right\rangle $, and all phases have nonzero $\langle
F_{z}\rangle $. As a result, the phase transitions between PPW and PW1 (PW2)
become crossovers (see Appendix).

Due to the hybridization between $|+\rangle $ and $|0\rangle $ for the $%
\left\vert \uparrow ^{\prime }\right\rangle $-band, $m_{z}$ would lower the
right minimum more significantly. Therefore the global minimum may change
from left to right as we increase $m_{z}$, which drives the phase
transitions from SS phase first to PW1 then to PW2 phases [see the left part
in Fig.~\ref{fig:mz_g0}(a)]. In addition, the transition from PW2 to PW1
occurs when the global band minimum is still the right one, which means that
the BEC prefers the high-energy local minimum at the left [as schematically
shown in the inset of Fig.~\ref{fig:mz_g0}(b)], where the hybridization
between $|+\rangle $ and $|0\rangle $ leads to a lower interaction energy
from $g_{2}n_{0}(n_{\uparrow }+n_{\downarrow })$ that compensates the higher
single-particle energy.

\emph{Conclusions.---}In summary, we propose a scheme to realize a novel magnetic
superstripe phase through engineering
a spin-1/2 subsystem with tunable SO coupling in a spin-1 BEC. The
tunable SO coupling could be generalized to other Bose and Fermi cold atomic systems,
including Alkali-earth(like) atoms.
The system does not suffer heating issues and are robust against magnetic field fluctuations, making it
a promising platform to explore supersolid physics (e.g., the phase transition, non-trivial dynamics, roton spectrum).
More importantly,
the superstripe phase has magnetic crystalline structure with
a high visibility and long tunable period that can be directly detected by real-space imaging.
Our scheme not only opens the
possibility for exploring novel physics with tunable SO coupling; but also
paves the way for conclusive (real-space) observation and exploration of long-sought
supersolid phases in experiments.


\textbf{Acknowledgements}: We thank P. Engels for helpful discussion. This
work is supported by AFOSR (FA9550-16-1-0387), NSF (PHY-1505496), and ARO
(W911NF-17-1-0128).

\newpage
\newpage
\begin{widetext}
\section*{Appendix}

\setcounter{figure}{0} \renewcommand{\thefigure}{A\arabic{figure}} %
\setcounter{equation}{0} \renewcommand{\theequation}{A\arabic{equation}}

\textbf{\emph{Other experimental configurations for generating tunable
spin-orbit (SO) coupling.---}}Our scheme for tunable 1D SO coupling only relies on the existence of three
hyperfine ground states that can be coupled with each other, therefore it
can be applied to other alkali (e.g., potassium) and alkaline-earth(-like)
atoms (e.g., strontium, ytterbium). The corresponding laser configurations
could be slightly different.

For instance, for fermionic $^{40}$K, we can choose $\left\vert \downarrow
\right\rangle =|F=\frac{7}{2},m_{F}=\frac{3}{2}\rangle $, $\left\vert
\uparrow \right\rangle =|F=\frac{9}{2},m_{F}=\frac{3}{2}\rangle $ and $%
\left\vert 0\right\rangle =|F=\frac{9}{2},m_{F}=\frac{1}{2}\rangle $ as the
spin states~\cite{huang2016experimental} [Fig.~\ref{fig:S0}(a)] with the
same laser configuration as that in Fig.~1(a) in the main text. The phases
of these Raman lasers are irrelevant because they can be gauged out in the
definition of spin states. The coupling $\delta $ between $\left\vert
\downarrow \right\rangle $) and $\left\vert \uparrow \right\rangle $ is
realized directly by the Raman lasers $\omega _{\uparrow }$ and $\omega
_{\downarrow }$~\cite{huang2016experimental}, and a positive $\delta >0$ can
be obtained using Raman lasers between D1 and D2 lines (no need for
microwave fields). Notice that for $^{87}$Rb in the $F=1$ manifold in the
main text, the two-photon microwave transition is used to couple $%
|F=1,m_{F}=1\rangle $ ($\left\vert \downarrow \right\rangle $) and $%
|F=1,m_{F}=-1\rangle $ ($\left\vert \uparrow \right\rangle $) because the
corresponding Raman coupling need be near-resonance, yielding significant
heating.

\begin{figure}[b]
\includegraphics[width=0.6\linewidth]{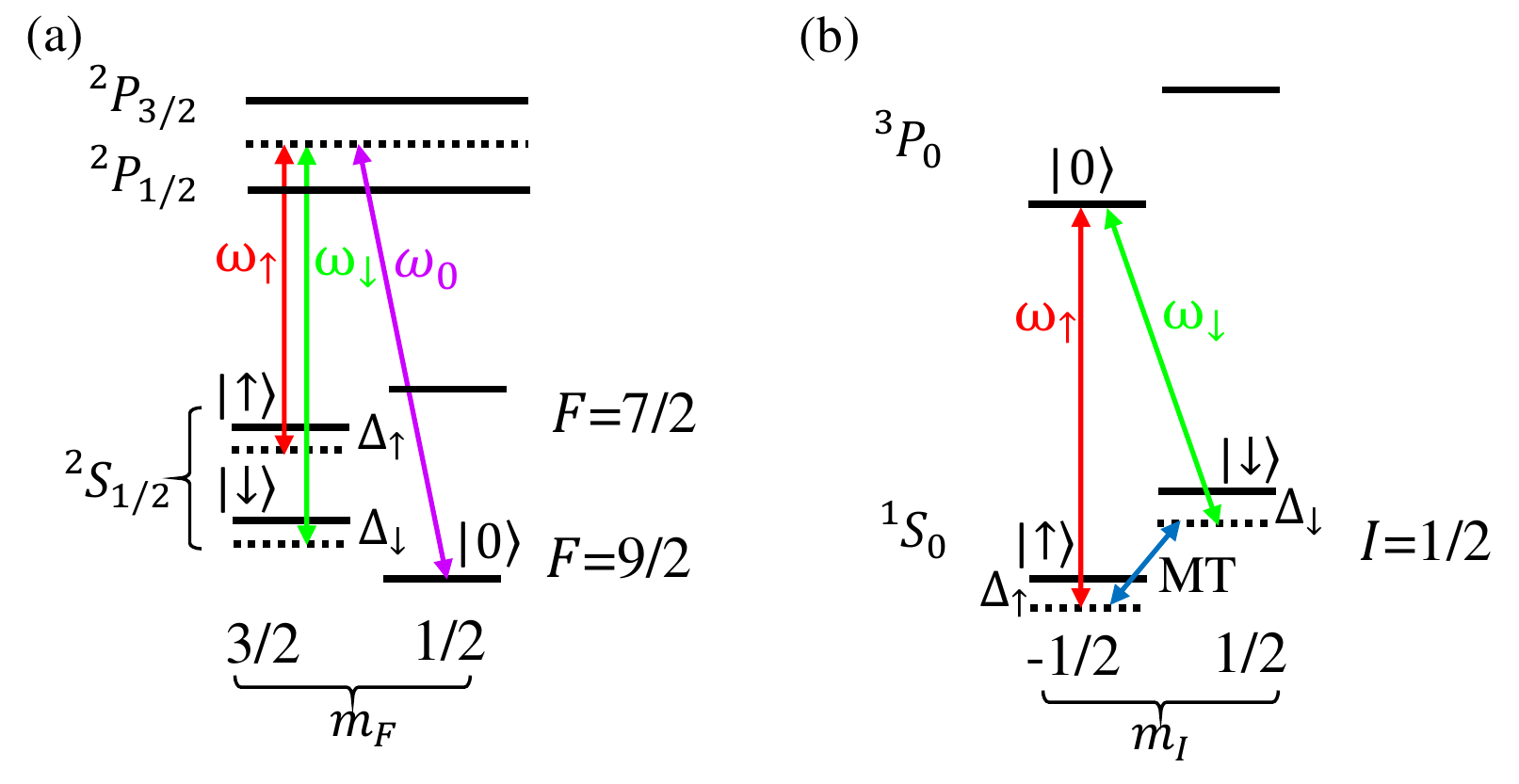}
\caption{(a) Energy levels and Raman transitions to generate tunable SO
coupling for $^{40}$K. (b) Energy levels and clock (microwave) transitions
to generate tunable SO coupling for $^{171}$Yb.}
\label{fig:S0}
\end{figure}

For the fermionic alkaline-earth(-like) atoms (e.g., $^{87}$Sr, $^{171}$Yb, $^{173}$Yb)~\cite{PhysRevA.76.022510,
mancini2015observation, bromley2018dynamics, kolkowitz2017spin}, we can use
two nuclear spin states in the $^{1}$S$_{0}$ manifold and one nuclear spin
state in the $^{3}$P$_{0}$ manifold to represent a spin-1 system [see Fig.~%
\ref{fig:S0}(b)]. Instead of a Raman process, the coupling between $%
\left\vert 0\right\rangle $ and $\left\vert \uparrow ,\downarrow
\right\rangle $ is realized by the one-photon Rabi transition (i.e., the
clock transition). The laser setup is similar as that in Fig.~1 in the main
text, except that only two laser beams $\omega _{\uparrow }$ and $\omega
_{\downarrow }$ are needed, whose polarizations are rotated by $\pi /4$ with
respect to $z$-direction. These lasers can generate both $\pi $- and $\sigma
^{-}$-clock transitions between $\left\vert 0\right\rangle $ and $\left\vert
\uparrow ,\downarrow \right\rangle $. A one-photon microwave transition is
also needed to achieve the coupling between $\left\vert \uparrow
\right\rangle $ and $\left\vert \downarrow \right\rangle $. To obtain a
positive $\delta $, the phase of the microwave field is locked to the same
value as the phase difference between two clock lasers. The couplings with
other nuclear spin states are suppressed due to the different Zeeman
splitting and dipole potential~\cite{PhysRevA.76.022510,
mancini2015observation}.
Similar spin-orbit coupling schemes can also be applied to fermionic species that are not considered in this work.

\textbf{\emph{Variational energy functional.---}}In the basis $\Psi =(\psi _{\uparrow },\psi _{0},\psi _{\downarrow })^{T}$,
the interaction energy in the laboratory frame is 
\begin{eqnarray}
\varepsilon _{\text{int}}\ &=&\frac{1}{V}\int dx\left[ \frac{g_{0}}{2}n^{2}+%
\frac{g_{0}}{2}(\Psi ^{\dag }\mathbf{F}\Psi )^{2}\right]  \nonumber \\
&=&\frac{1}{V}\int dx\left[ \frac{g_{0}}{2}n^{2}+g_{2}n_{0}(n_{\uparrow
}+n_{\downarrow })+\frac{g_{2}}{2}(n_{\uparrow }-n_{\downarrow })^{2}\right]
+\frac{1}{V}\int dx2g_{2}\Re \lbrack \psi _{\uparrow }\psi _{\downarrow
}\psi _{0}^{\ast }\psi _{0}^{\ast }],  \label{eq:inter-energydensity_a}
\end{eqnarray}%
%
where $\mathbf{F}=(F_{x},F_{y},F_{z})$. After the unitary transformation $%
U=\exp (i2k_{\text{R}}x)\left\vert 0\right\rangle \left\langle 0\right\vert
$ to the quasi-momentum frame, the above equation is unchanged except that
the last term becomes
\begin{equation}
\frac{1}{V}\int dx2g_{2}\Re \lbrack \psi _{\uparrow }\psi _{\downarrow }\psi
_{0}^{\ast }\psi _{0}^{\ast }\times \exp (4ik_{R}x)],
\end{equation}%
which is nonzero only when the state is a superposition of two plane waves
with momentum separation $2k_{R}$. Here we focus on the case where the
momentum separation is much smaller than $2k_{R}$, therefore this term
becomes zero and we obtain the interaction energy Eq.~(2) in the main text.

Using the variational ansatz
\begin{eqnarray}
\Psi=\sqrt{\bar{n}}|c_1|\left(
\begin{array}{c}
\cos(\theta_1)\cos(\phi_1) \\
-\sin(\theta_1) \\
\cos(\theta_1)\sin(\phi_1) \\
\end{array}
\right)e^{ik_1x} +\sqrt{\bar{n}}|c_2|\left(
\begin{array}{c}
\cos(\theta_2)\cos(\phi_2) \\
-\sin(\theta_2) \\
\cos(\theta_2)\sin(\phi_2) \\
\end{array}
\right)e^{ik_2x+i\alpha},
\end{eqnarray}
we obtain the single particle energy density
\begin{eqnarray}
\varepsilon_\text{0}&=& \frac{1}{V}\int dx\Psi ^{\dag }H_{0}\Psi  \nonumber
\\
&=&\bar{n}\sum_{i}|c_i|^2\left\{k_i^2+\sin(2\theta_i)\sin(\phi_i+\frac{\pi}{4%
})+[\Delta+4-4k_i+\delta\sin(2\phi_i)+m_z\cos(2\phi_i)]\cos^2(\theta_i)%
\right\},
\end{eqnarray}
and the interaction energy density
\begin{eqnarray}
\varepsilon_\text{int}&=&\bar{n}\frac{g_0\bar{n}}{2}\left%
\{1+2|c_1|^2|c_2|^2[\sin(\theta_1)\sin(\theta_2)+\cos(\theta_1)\cos(%
\theta_2)\cos(\phi_1-\phi_2)]^2\right\}  \nonumber \\
& &+\bar{n}\frac{g_2\bar{n}}{2} \bigg\{2|c_1c_2|^2\left[\cos^2(\theta_1)%
\cos^2(\theta_2)\cos^2(\phi_1+\phi_2) +
\sin(2\theta_1)\sin(2\theta_2)\cos(\phi_1-\phi_2)\right]  \nonumber \\
& & +\left[\sum\nolimits_i|c_i|^2\cos^2(\theta_i)\cos(2\phi_i)\right]^2 +2%
\left[\sum\nolimits_i|c_i|^2\sin^2(\theta_i)\right]\left[\sum%
\nolimits_i|c_i|^2\cos^2(\theta_i)\right]\bigg\},
\end{eqnarray}
with the total energy density given by $\varepsilon_\text{tot}%
=\varepsilon_0+\varepsilon_\text{int}$.

\begin{figure}[b]
\includegraphics[width=0.6\linewidth]{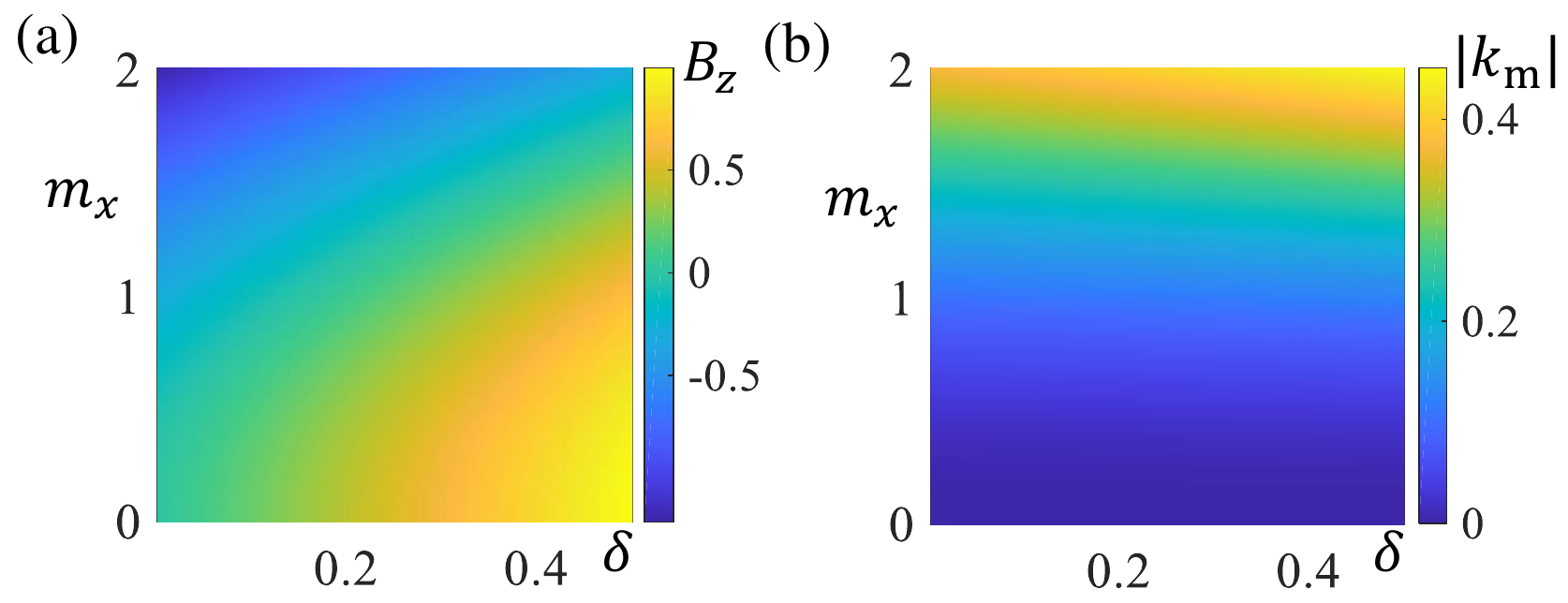}
\caption{(a) and (b) The dependence of $B_{z}$ and $k_{\text{m}}$ on the
Raman and microwave couplings $m_{x}$, $\protect\delta $. The parameters are
$\Delta =-2$, $m_{x}=1$, $\protect\delta =0.18$, $m_{z}=0$.}
\label{fig:S1}
\end{figure}

\begin{figure}[t]
\includegraphics[width=0.7\linewidth]{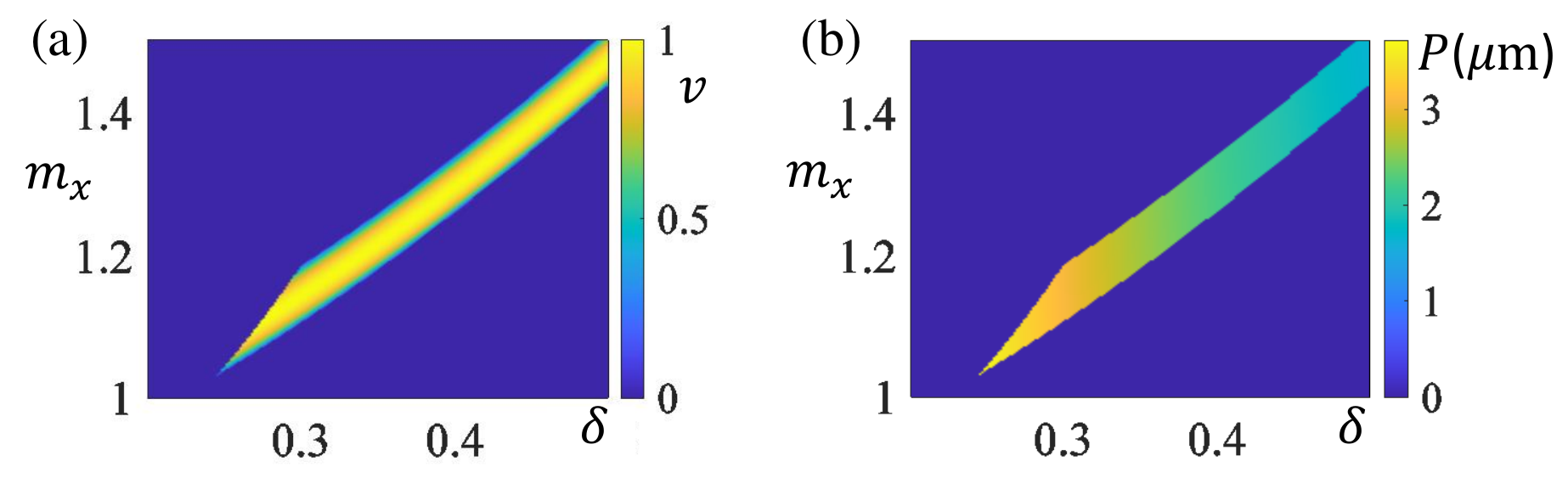}
\caption{(a) and (b) The corresponding visibility ($v$) and period (P) of
the spin density modulations in the SS phase shown in Fig.~1(a) in the main
text, with $v\equiv \frac{\max (n_{\uparrow ,\downarrow })-\min (n_{\uparrow
,\downarrow })}{\max (n_{\uparrow ,\downarrow })+\min (n_{\uparrow
,\downarrow })}$. We set $P=v=0$ in the (polarized) plane-wave phases, and
the maximum period is the SS phase is about 3.8 $\protect\mu $m.}
\label{fig:S2}
\end{figure}

\begin{figure}[b]
\includegraphics[width=0.6\linewidth]{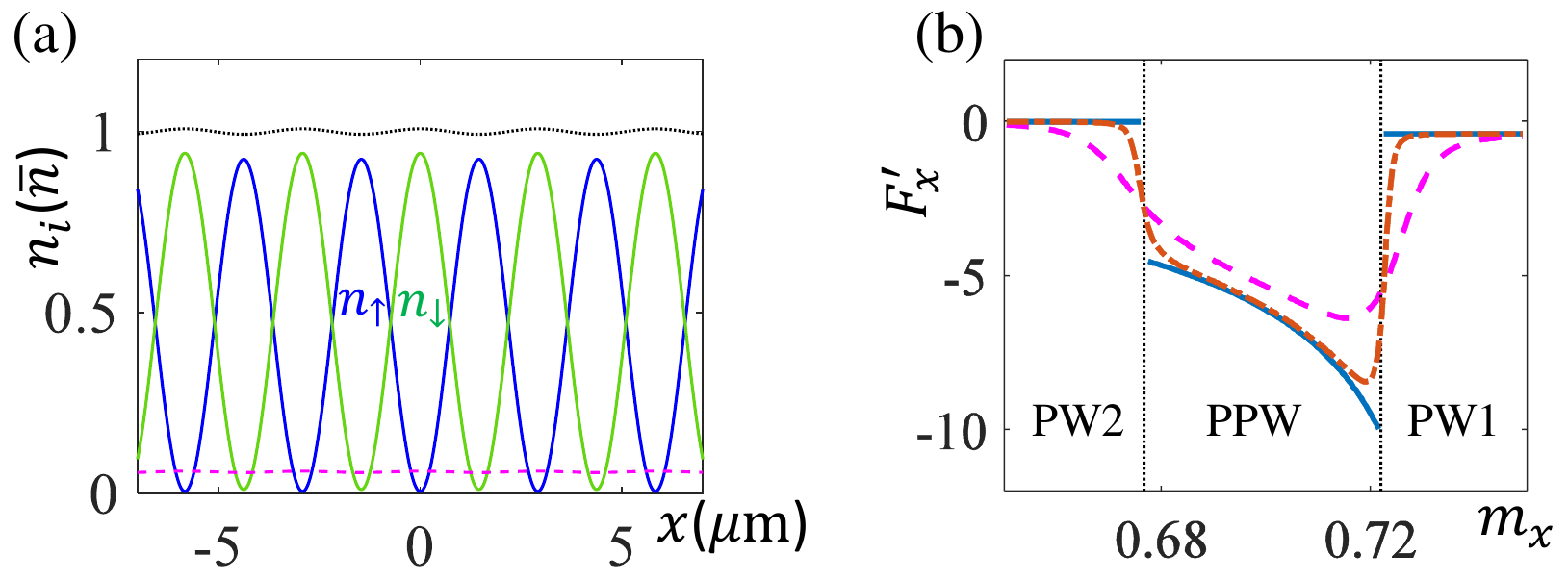}
\caption{(a) Spin density modulation in the SS phase with $\protect\delta %
=0.3$ and $m_{z}=0.8g_{2}\bar{n}$. Both the total density and spin densities
have slight periodic modulations. Other parameters are the same as Fig.~4(a)
in the main text. (b) Change from second-order phase transitions to
crossovers due to finite Zeeman field $m_{z}$. From the first order
derivative of $F_{x}$ over $m_{x}$ (which equals to the second-order
derivative of $\protect\varepsilon _{\text{tot}}$ over $m_{x}$ due to the
Hellmann-Feynman theorem, i.e., $F_{x}^{\prime }=\protect\varepsilon _{\text{%
tot}}^{\prime \prime }$), we see that that The PPW-PW1 and PPW-PW2
boundaries change from second-order boundaries with $m_{z}=0$ (blue solid
line) to crossover boundaries with $m_{z}\neq 0$ (red dash-dotted line for $%
m_{z}=10^{-4}$ and purple dashed line for $m_{z}=10^{-3}$). Other parameters
are $\Delta =-1$, $\protect\delta =0.1$.}
\label{fig:S3}
\end{figure}

\textbf{\emph{Some details about tunable SO coupling and superstripe phase.---}}
As we discussed in the main text, the low energy dynamics are characterized
by an effective spin-1/2 system with tunable SO coupling, with an effective
Hamiltonian (in the basis $\{\left\vert \uparrow ^{\prime }\right\rangle
,\left\vert \downarrow ^{\prime }\right\rangle \}$)
\begin{equation}
H_{\text{eff}}=\left[
\begin{array}{cc}
\eta (k-k_{\text{m}})^{2} & 0 \\
0 & k^{2}%
\end{array}%
\right] +B_{z}\sigma _{z}+B_{x}\sigma _{x}.
\end{equation}%
The transverse field $B_{x}$ is approximately given by the Zeeman field $%
m_{z}$, while the longitudinal field $B_{z}$ and the SO coupling strength $%
k_{\text{m}}$ can be tuned by varying Raman and microwave coupling strengths
$m_{x}$ and $\delta $. In Fig.~\ref{fig:S1}, we plot the dependence of $B_{z}
$ and $k_{\text{m}}$ on $m_{x}$ and $\delta $.

Due to the tunability of the SO coupling, we can obtain superstripe phases (SS) with a
tunable and long period, and the effects of density interaction can be
suppressed by a weak or vanishing $B_{x}$, which leads to a high-visibility
spin superstripe phase favored by the spin interaction. In Fig.~\ref{fig:S2}, we
plot the period and visibility of the superstripe phase in Fig.~1(a) in the main
text.

We notice that the SO coupling is written in the basis $\left\vert \uparrow
^{\prime }\right\rangle $ and $\left\vert \downarrow ^{\prime }\right\rangle
$ (which are approximately given by $\left\vert \uparrow ^{\prime
}\right\rangle \simeq |+\rangle =\frac{1}{\sqrt{2}}(\left\vert \uparrow
\right\rangle +\left\vert \downarrow \right\rangle )$ and $\left\vert
\downarrow ^{\prime }\right\rangle =|-\rangle =\frac{1}{\sqrt{2}}(\left\vert
\uparrow \right\rangle -\left\vert \downarrow \right\rangle )$), while the
spin density modulation is formed in a different basis $\left\vert \uparrow
\right\rangle $ and $\left\vert \downarrow \right\rangle $. A natural
question to ask is whether high-visibility spin density modulation can be
obtained in the basis $|+\rangle $ and $|-\rangle $ for conventional SO
coupling scheme in the basis $\left\vert \uparrow \right\rangle $ and $%
\left\vert \downarrow \right\rangle $. The answer is no and the reason is
illustrated below. In our scheme, only state $|0\rangle $ is transformed to
the quasi-momentum frame, and states $\left\vert \uparrow \right\rangle $
and $\left\vert \downarrow \right\rangle $ are associated with atomic
mechanical momentum, therefore the plane-wave superposition of $|+\rangle $
and $|-\rangle $ at different momenta gives rise to spin density modulation
in the laboratory frame, with period directly determined by the momentum
difference between two plane waves. While for conventional SO coupling
scheme in the basis $\left\vert \uparrow \right\rangle $ and $\left\vert
\downarrow \right\rangle $, both states $\left\vert \uparrow \right\rangle $
and $\left\vert \downarrow \right\rangle $ are transformed to the
quasi-momentum frame, and the superstripe state in the quasi-momentum frame is~%
\cite{li2012quantum}
\begin{equation}
(\left\vert \uparrow \right\rangle +\epsilon \left\vert \downarrow
\right\rangle )e^{-ikx}+(\left\vert \downarrow \right\rangle +\epsilon
\left\vert \uparrow \right\rangle )e^{ikx}.
\end{equation}%
In the ideal case, we may have $k\simeq k_{\text{R}}(1-g_{2}/2g_{0})$ and $\epsilon \simeq
\sqrt{g_{2}/2g_{0}}$. After transforming back to the laboratory frame, the
above state in the basis $|\pm \rangle $ can be written as
\begin{equation}
\lbrack \cos (kx-k_{\text{R}}x)+\epsilon \cos (kx+k_{\text{R}}x)]|+\rangle
+[\sin (kx-k_{\text{R}}x)+\epsilon \sin (kx+k_{\text{R}}x)]|-\rangle .
\end{equation}%
Without loss of generality, we consider the $|+\rangle $ state, where
$\epsilon \cos (kx+k_{\text{R}}x)$ gives a short-period modulation ($\sim0.4\mu$m) with
a low visibility $\sim \sqrt{g_{2}/2g_{0}}$ that is around $5\%$ for typical
parameters of $^{87}$Rb, while $\cos (kx-k_{\text{R}}x)$ gives an extremely
long-period modulation around $300\mu $m that is invisible for typical BEC
cloud size (less than 100$\mu $m).

In the presence of $m_{z}$ (i.e., $B_x\neq0$), the spin states at two band minima are no longer
orthogonal, and the SS phase now possesses both spin and total density
modulations, as shown in Fig.~\ref{fig:S3}(a). The Zeeman field $m_{z}$
breaks the $Z_{2}$ symmetry between $\left\vert \uparrow \right\rangle $ and
$\left\vert \downarrow \right\rangle $, and all phases now have nonzero $%
\langle F_{z}\rangle $. The phase transitions between PPW and PW1 (PW2)
become crossovers, as confirmed by our numerical results of the derivative
of the ground-state energy [see Fig.~\ref{fig:S3}(b)].



\end{widetext}




\end{document}